# IMPLEMENTATION OF SERVICE-ORIENTED ARCHITECTURE FOR eWALLET SYSTEM FOR CASHLESS TRANSACTIONS IN THE Democratic Republic of Congo

*Patrick Mukala*
*Department of Software engineering*
*Faculty of Information and Communication Technology, Tshwane University of Technology, Kinshasa, South Africa*
mukalapm@tut.ac.za

## 1. ABSTRACT

The Democratic Republic of Congo is a sleeping giant at the heart of Africa. Though endowed with incredible natural resources spanning from significant reserves of gold, diamond, coltan, copper to vast African forests and impressive water resources, it is still considered as one of the poorest countries on the planet. For a long time, its economy has had nothing to show for all these resources as the Congolese people continue to live in extreme poverty. To survive, the Congolese people engage in a range of activities to generate income and most families rely on "solidarity grants" from family members and relatives that either work or conduct a personal business. Most of these families remain unbanked. Due to the absence of reliable and private financial institutions, almost 90% of all financial transactions are cash-based. However, as the number of banks increases throughout the past few years, there is a possibility to encourage and perform cashless transactions. In this paper, we demonstrate an implementation of a Service-Oriented Architecture approach in the design of a eWallet system. A typical scenario has been presented to highlight the major components of the system to be implemented as contextualized in the ecosystem. A series of models, from the business processes to the technical architecture and service model for the system, has been developed to represent every step of the solution.

The paper is structured as follows: In section 1, we give a brief introduction of the paper, section 2 gives a generic description and contextualization of eWallet, section 3 gives a description of the scenario as chosen for this case study, section 4 describes the ecosystem for the scenario, section 5

## 2. INTRODUCTION

For a lot of reasons either economic or political, the vast part of Africa still does not have access to proper banking services. Most African people remain unbanked. According to a study conducted by the First National Bank (FNB), more than 13 million adults in South Africa still remain unbanked, with the biggest percentage of these adults based in remote or rural towns. A similar research was conducted in Botswana by Finscope in 2009 survey that revealed that more than half of the adult population in Botswana remains unbanked. It is the same situation in the Democratic Republic of Congo.

Third largest country in Africa, the Democratic Republic of Congo span on 2,345,410 square km sharing borders nine countries. Despite its vast mineral resources that earned it the nickname "geological scandal" , its population still lives in hopeless poverty (Ngoma, 2010). While the country has been crippled for more than a decade by war, its population survives mostly on small scale retailing (personal business) due to high rate of unemployment. This state of events has resulted in a situation where the majority of the Congolese people rely on grants to survive. All these transactions including many more transactions in both the public and private sectors as well as between individuals or organizations are cash transactions.

To facilitate some of these transactions for remote distances, there are numbers of money transfer agencies across the country that provide cash-based transactions. Some of these agencies include Money Gram, moneytrans, Western Union and Soficom (the most popular), Solitaire, Kin Service express and so forth. However, this does not always provide the best means to move currency as inherent risks involve loss, robbery, mismanagement etc. But as the country's banking sector has started to slowly emerge with the establishment and expansion of banks such as BIAC,BIC, EcoBank, Afriland bank, ProCredit Bank, Trust Merchant bank (TMB), Standard bank, SofiCom Bank to name but a few, the idea of electronic financial transactions has started to gain access in the mind of the few banked DRC population. However, the majority of the DRC population remains unbanked. Hence, this situation has made monetary transactions between the affected individuals and the rest of the world difficult and almost non-existent. The important question we try to tackle in this paper is : How do we then facilitate the link between the banked population and the unbanked population to in order to facilitate financial transactions?

However, with the expansion of telecommunications (presence of big companies such as Vodacom, Airtel, Tigo and CCT) and their availability even in DRC remote areas, technology has made it possible to overcome this challenge. A new concept called eWallet can be adopted to introduce mobile money or electronic money transfer services.

Mobile money is one of the fastest growing sectors in the financial services environment - with many people opting for an electronic wallet instead of a leather one. Mobile money transfer services have become the most convenient way of sending money, as with cellphones one is able to reach anyone from any part of the country. This service (cell phone banking) is already available in most of the country's' banks and many people already own mobile phones capable of carry out this service. This service offers numerous benefits including allowing financial transactions among those classified as "unbanked" to bridge the gap between them and the so called economic active "banked" population.

Furthermore, through cell phone and Internet banking, eWallet can be activated to cater for various financial needs. According to Newswire Association (2011), eWallet will enable multi-national companies to pay its members or employees in foreign countries through a web-based portal in several currencies. Account-holders will be able to transfer money from their online account or by cell-phone to bank accounts or prepaid debit cards. In addition, the new platform will significantly increase revenue and will replace the existing vendor platform relationships (Newswire Association LLC, 2011).

In this article, we explore the concept of eWallet and its relevance to the needs of both the unbanked and banked population in the DRC. A context on eWallet is given followed by a series of models presented on the basis of a given scenario to document the implementation of the technology.

## 3.  eWALLET: DEFINITION AND CONTEXT

In a broad context, eWallet can be described as a service that allows financial and monetary transactions for anyone with a valid cell phone number within the confines of a specific environment. When activated, it allows the account holder to withdraw money from an ATM whether the recipient does or does not hold a bank account. The money can also be used to buy airtime, pay for goods and services online, and to transfer to another eWallet (National Traveller MABS, 2011).

The service requires that the sender have an active bank account to transfer money using mobile phone to the recipient who doesn't have to have a bank account.  The receiver gets an automated short service message (SMS) informing them about the amount sent to them, and can claim the amount using a temporary pin at the nearest Auto Teller Machine (ATM ) or pay goods and services online and or with participating merchandise.
Hence, eWallet can be perceived nowadays as a convenient, quick and safe way to perform financial transactions compared to the traditional channels of sending money which are sometimes informal, costly, insecure and unreliable.

## 4.  UNDERSTANDING THE PROBLEM

### 4.1  SCENARIO

In order to illustrate the potential of eWallet, we consider a scenario in a South African context. This scenario, taken as the main case study, will be used to design a series of typical models to document the service. The scenario involves family members with Mr. Kayembe Ka Tshitupa  being the only banked member of the family. Since the rest of the family members rely on Mr. Kayembe Ka Tshitupa , geographical constraints call for a new and quick way of sending money to the family. The scenario is detailed as follows.

Mr. Kayembe Ka Tshitupa originates from Kasai Occidental (Western Kasai) province, and is working as a Teacher in Kinshasa. His parents live in Kananga from his native province. Mr. kayembe has a lot of family members that depend on his salary. Every month, he sends money to his parents, his brothers for their fees and other expenses and any other relative in need. These transactions cost him extra money as the different agencies he makes use of charge him up to 10 % of the initial amount being sent. As the DRC government has introduced a new payment method for all public officials including teachers through local private banks, He is now a bank account holder.

However, Mr. Kayembe Ka Tshitupa  has a challenge on how to send the money because neither his parents nor his other relatives have a bank account. Hence, the default resolution to get money to his parents will be to use physical means by either meeting his them in person, or sending the money with a third party through an agency or an acquaintance ( which is not always guaranteed that the money will reach the destination). This means he will have to travel each and every month from Kinshasa to Kananga to deliver the money to his parents…, But this incurs extra expenses as it requires spending too much money on transport and travelling to Kananga every month.

To solve the dilemma, Mr. Kayembe Ka Tshitupa has decided to buy cellular phones for his wife and parents to reduce spending on transport and long travelling. He can create and activate the eWallet service for himself, the wife and his parents. This way, he can use his bank account to recharge his eWallet account and from this account, he can transfer any available amount of money to his wife's and parents' respective eWallet accounts. Once the wife has an eWallet account, she has the ability to perform an inter account transfer with anybody else holding an eWallet account.

To send the money, Mr. Kayembe Ka Tshitupa will initially use a bank branch, ATM, cellphone (his own eWallet account) or use the Internet to perform the transactions. When sending the money, all he needs is his wife's cellphone numbers. The wife will receive an appropriate SMS containing a temporary secret code or PIN number and the amount transferred. Upon receiving the SMS from eWallet System, she can transfer certain amount to the parents but she must have their cell phone numbers. The cellphone number can from any cellphone provider of choice, e.g. MTN, Vodacom, Cell C or Telkom.

The wife will, in return, decide on whether to withdraw the whole amount or half of the money from any ATM, or purchase from participating Seller. When she purchases goods, she will be required to enter the cellphone number of Seller. The Seller will have to enter the temporary pin in order to verify the transaction. Mr. Kayembe Ka Tshitupa 's wife can now receive goods without having cash on hand.

Implementing this service, one will see that eWallet is a convenient way to overcome the challenges previously experienced by the family. Despite the lack of bank accounts from the rest of the family members, Mr. Kayembe Ka Tshitupa is able to send money and procure goods for the family on a regular basis thanks to eWallet and hence saving more than spending too much money on cash-based transactions.

## 4.2 ECOSYSTEM

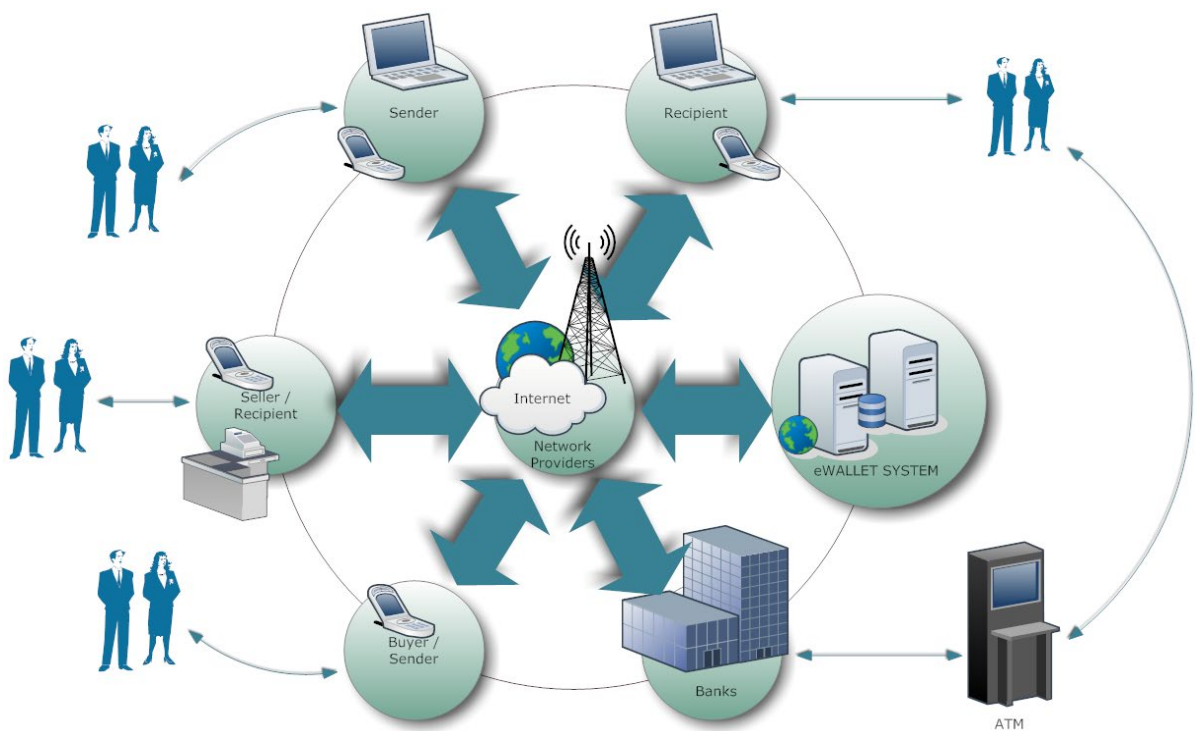

## Stakeholders in the Ecosystem

| ACTORS/STAKEHOLDERS | ROLE & RESPONSIBILITIES |
|---|---|
| 1. Sender | The person initiating money transfer from his account to someone else either for the purpose of transferring money or paying for a service and/or goods |
| 2. Cellphone | Device used to interact with the system. The cellphone becomes indeed, the sender wallet. |
| 3. Network provider | Links between the sender and the system. |
| 4. eWallet System | This is the eWallet system made of several databases to keep records of every transactions and to update bank accounts accordingly |
| 5. Bank | Provide link between the sender bank account and the system |
| 6. ATM | Device allowing the receiver to withdraw cash that has been sent to him by the sender |
| 7. Seller / Store | Any store from which an eWallet account holder can purchase from. The Seller becomes the recipient and the buyer the sender |
| 8. Recipient | The person to whom money was sent to. |

The eWallet system outsources services from different banks and network providers. A special service command number is allocated to the system (e.g. "#555*").the prerequisite for sending money using the service is to signup, upon which the system will allocate an account to the sender's cellphone number and a security pin number. (This information is sent to the user via SMS)

The sender dials "#555*" from his cellphone, the networks provider redirects the request to the eWallet system that replies with a prompt menu.
1. Transfer money
2. Withdraw money
3. Change pin number
4. Check your balance

To send money, the sender replies with 1
(Two Options: transfer money to your eWallet, transfer money to someone else)
The system then prompts the sender with his security pin number, the system prompt for the amount to be transferred. If the user has enough funds in his eWallet account, the system asks him from where the amount should be withdrawn.
1. Bank account

2. eWallet account

Otherwise, the amount would be withdrawn from his bank account.

The system prompts for the cellular number of the receiver and validates it on the local network providers' database. The system transfer the money to a temporary account (the eWallet bank account).if the transfer is successful, the system sends a "transaction successful" message to the user, and otherwise an appropriate error message is sent

If the transaction is successful, the receiver gets an SMS with the details of the transaction and instruction on how to retrieve the money ("dial #555*).

The receiver dials #555*, several options including"transfer money to a bank account", "transfer money to your eWallet" and"transfer money to someone else" are presented.

If the receiver has an account with the eWallet system, he is prompted either to withdraw the money, or to save the money into his account.

E.g. You have an incoming R550
    1. Withdraw the money
    2. Save it into your account

If the receiver does not have an account he will be prompted for the following:
    You have an incoming R550
    1. Withdraw the money
    2. Create an account to save your money

If the receiver chooses to withdraw the money an access code is generated to allow him to withdraw the money from any ATM.

## 4.3 VALUE PROPOSITION FOR ACTIVE PARTICIPANTS

The main participants for this scenario are the sender, receiver, service providers (banks, networks). For each of these participants, the benefits incurred in using the system depend on the services they need from the system.

Hence, for the sender, the system is valuable to him in the sense that it helps reduce costs associated with the physical way of sending money (either standing in long queues in a bank to deposit money for receivers with a bank account or driving for hours to reach a receiver that does not hold a bank account).

For the receiver, the system offers him/her the luxury of receiving money even though they don't have a bank account. The system is also time efficient for both the sender and the receiver.

Lastly, for the providers (bank, networks), the system will benefit them in expanding their brand and their services. Also, it will generate more revenue because of the fees charged on services provided during transactions.

## 5. PROCESSES

The business processes in the system depict the core functionalities of the application. To illustrate all the possible business processes, we start with a trigger which is commonly a need (need to start a transaction or create an account).

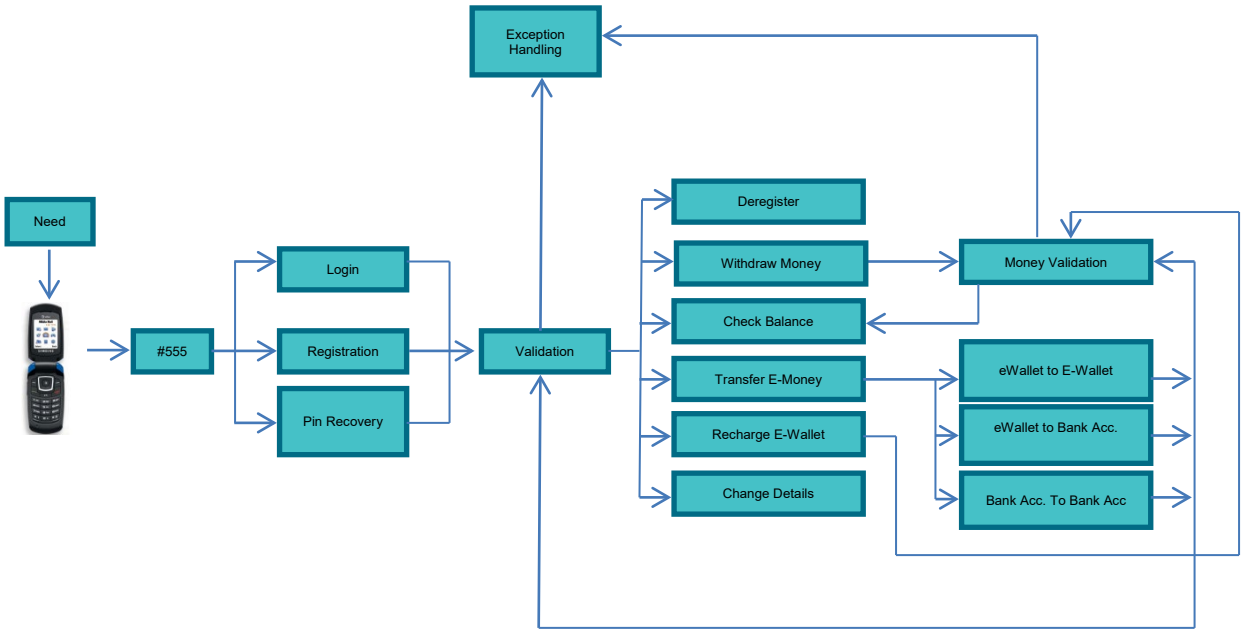

## A. LOGIN:

The logon process is twofold: from the subscriber's cellphone and from the Internet (online platform).
From a cellphone, the user is prompted for a secret pin, after authentication from cellular companies, access to main menu is granted.
When a user requests access from an online platform, he/she is asked for a username which is by default their cellphone number, then the password is requested before the User Data validation process kicks in, if there is a problem an exception is thrown (Error message: Invalid login details) if necessary and update the database (Keep track of logs).

## B. REGISTRATION:

It is a prerequisite that the client complete an application form either manually or online. The eWallet system will then confirm the client banking information with the Bank system. If the information is confirmed, the client will be registered on the eWallet system client registration database. Otherwise the eWallet system will return an error requiring the client to complete the form again or cancel registration.

I. The eWallet system will allocated a unique Login ID and temporary password
II. A system generated Login ID and password will be send via SMS to your registered cellphone number.

## C. PIN NUMBER RETRIEVAL:
- Prompt Cellphone Number (only for internet users)
- Display /SMS Secret question
- Prompt an Answer to secret question
- Validate user input (Using the User Data Validation functionality)
- Throw an exception (Error message: Invalid Answer) / Display or SMS the secret Pin Number
- Update Databases (logs).

This can be graphically represented as follows:

## D. USER DATA VALIDATION:
- Validate user data against E-Wallet internal database
- Validate cellphone number against Network providers databases
- Validate Secret Answer in the E-Wallet internal database

- Validate Pin Number in the E-Wallet internal database
- Validate Bank Account against Bank providers databases
- Generate appropriate error messages

This can be graphically represented as follows:

### E. MONEY WITHDRAWAL:
- Prompt Amount
- Validate amount against internal database (Using the Check Balance functionality)
- Update Database / throw an exception (Error message: Not sufficient funds)
- Generate a Temporary Pin (That would be used to withdraw the amount at the nearest ATM).

This can be graphically represented as follows:

### F. BALANCE CHECK:
- Display or SMS the actual E-Wallet balance

This can be graphically represented as follows:

### G. E-WALLET TO E-WALLET TRANSFER:
- Prompt recipient Cellphone Number
- Validate Recipient Cellphone Number (Using the User Data Validation functionality)
- Prompt amount to be transferred
- Validate amount (Using the Check Balance functionality)

- Update Database + Display or SMS a transaction successful message to the sender + notification to the recipient / throw an exception (Error message: Not sufficient funds) + logs.

This can be graphically represented as follows:

### H. E-WALLET TO BANK ACCOUNT TRANSFER:
- Prompt Recipient Bank Account
- Validate Recipient Bank Account (Using the User Data Validation functionality)
- Prompt amount to be transferred
- Validate amount (Using the Check Balance functionality)
- Update Database + Display or SMS a transaction successful message to the sender + notification to the recipient / throw an exception (Error message: Not sufficient funds) + logs

This can be graphically represented as follows:

### I. BANK ACCOUNT TO BANK ACCOUNT TRANSFER:
- Validate that the sender has a Bank Account (Using the User Data Validation functionality)
- Throw an exception (Error message) if necessary
- Prompt Recipient Bank Account
- Validate Recipient Bank Account (Using the User Data Validation functionality)
- Prompt amount to be transferred
- Validate amount (Using the Check Balance functionality)
- Update Bank accounts
- Update Database + Display or SMS a transaction successful message to the sender + notification to the recipient / throw an exception (Error message: Not sufficient funds) + logs

This can be graphically represented as follows:

### J. RECHARGE E-WALLET (BANK ACCOUNT TO E-WALLET TRANSFER ):

- Prompt Amount
- Validate that the sender has a Bank Account (Using the User Data Validation functionality)
- Throw an exception (Error message) if necessary
- Validate amount (Using the Check Balance functionality)
- Update Bank accounts
- Update Database (Add amount) + Display or SMS a transaction successful message to the sender / throw an exception (Error message: Not sufficient funds) + logs

This can be graphically represented as follows:

### K. UPDATE / CHANGE DETAILS:
- From Internet:   Display current details
  Prompt for new details
  Validate new details (Using the User Data Validation functionality)
  Update database
  Successful / Unsuccessful message + reasons
- From cell phone:   Prompt for new details (Only Cellphone and Pin number can be change via this media)
  Validate new details (Using the User Data Validation functionality)
  Update database
  SMS Successful / Unsuccessful message + reasons.

This can be graphically represented as follows:

## 5. TECHNICAL ARCHITECTURE

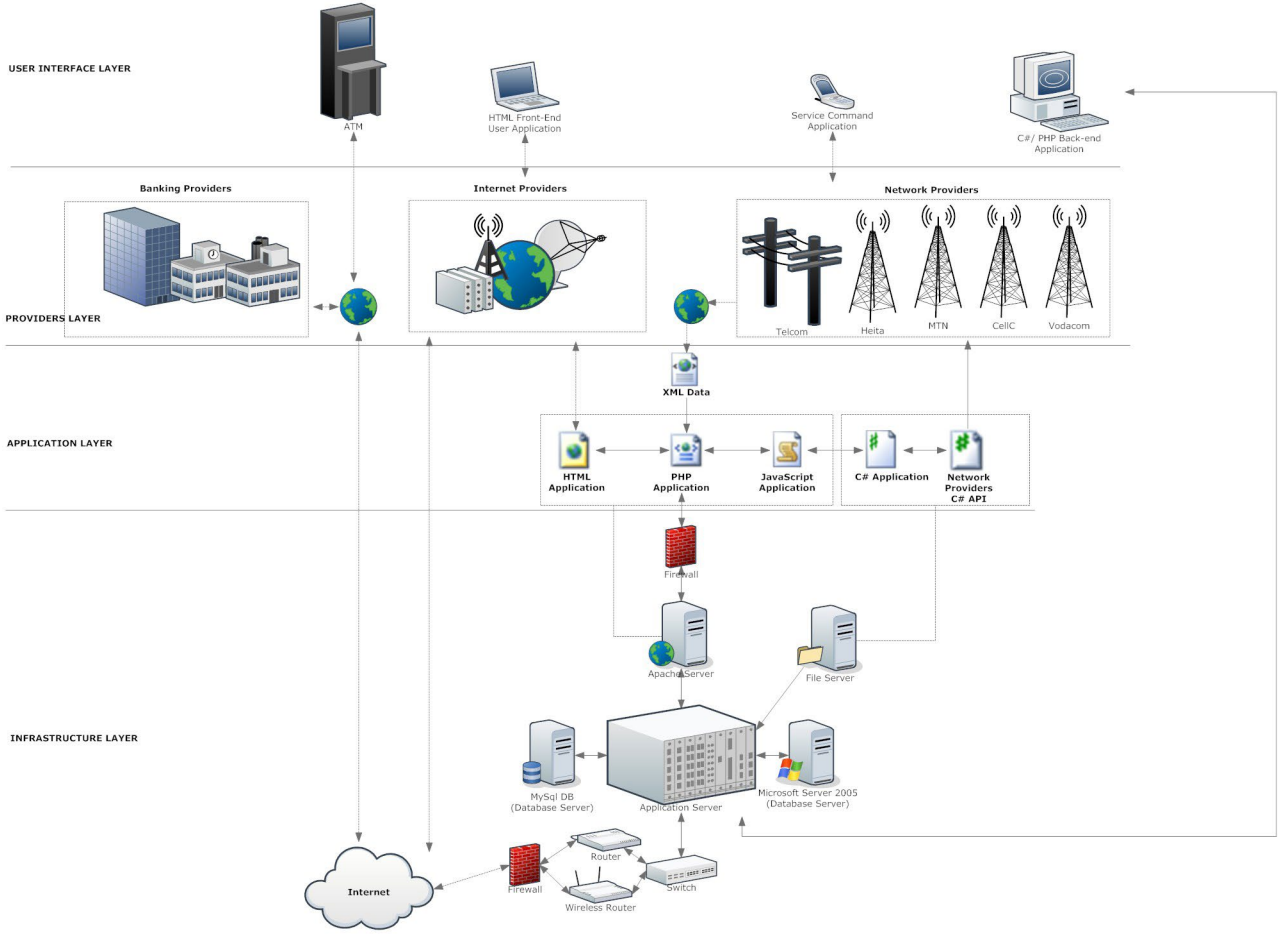

The technical architecture for the system as depicted in the above model is made up of four main layers namely the user-interface layer, the providers (carriers) layer, the application layer and the infrastructure layer.

A. **User-interface Layer:** This layer of the technical architecture represents the front-end of the system. It includes different devices needed to access the eWallet system. These devices include an Automatic Teller Machine (ATM), a cellular phone and a personal computer (desktop machine or laptop). Other devices range from Personal digital Assistant (PDAs), smart phones and tablet personal computers such as iPad and Samsung galaxy or any other portable devices that provide internet access to the user. For the system administrators, this layer provides a machine that connects directly to the back-end application for any administrative operations on the system.

B. **Providers' Layer:** This layer is composed of all the service providers that enable the communication between the user-interface and the application layer. These are Internet Service Providers (ISPs) that allow access to the system for users having access to the internet; telecommunication companies such as Vodacom, MTN and Cell C depending on the user's phone number; banks such as FNB, Standard Bank or Absa that host the main bank account to which all the eWallet account are connected and from which they are managed.

The services from these providers depend on each other and are monitored from the system itself in such a way that every provider is required to timely execute its allocated instructions to satisfy the user's request. A request from a cell phone for example needs processing from a dedicated telecommunication company (Cell C, MTN or Vodacom), which will validate the authenticity of the cell phone. An interlink between the cell phone and an eWallet Account requires a further verifications between the cell phone company and a specific bank

C.  **Application Layer:** In the application layer, there is a series of sub-systems dedicated to different functionalities of the system. This means that one subset of the system is dedicated to interact and monitor a specific service provider; another subset of the system validates data from all the transactions occurring on the system and so forth ; another one monitors and coordinates the interaction between all the other subsets. XML, HTML, PHP and JavaScript are the different scripting languages used to design the system with all its subsystems. The interaction between all the other layers of the technical architecture is specified and managed from the application layer.

D.  **Infrastructure Layer:** This layer consists of several technical devices that house the details of the application. The devices range from servers to communication routers in order to facilitate the connection between user's access devices to Internet providers and banks. The layer serves as the actual back-end of the system. A dedicated server (application server) stores the codes and scripts part of the system, another server stores the data repository. Additional servers include a server for eWallet accounts, another server for filing etc. A firewall for security purposes, as connection also occurs via internet and finally a switch to coordinate the different connections.

7. **SERVICE MODEL**

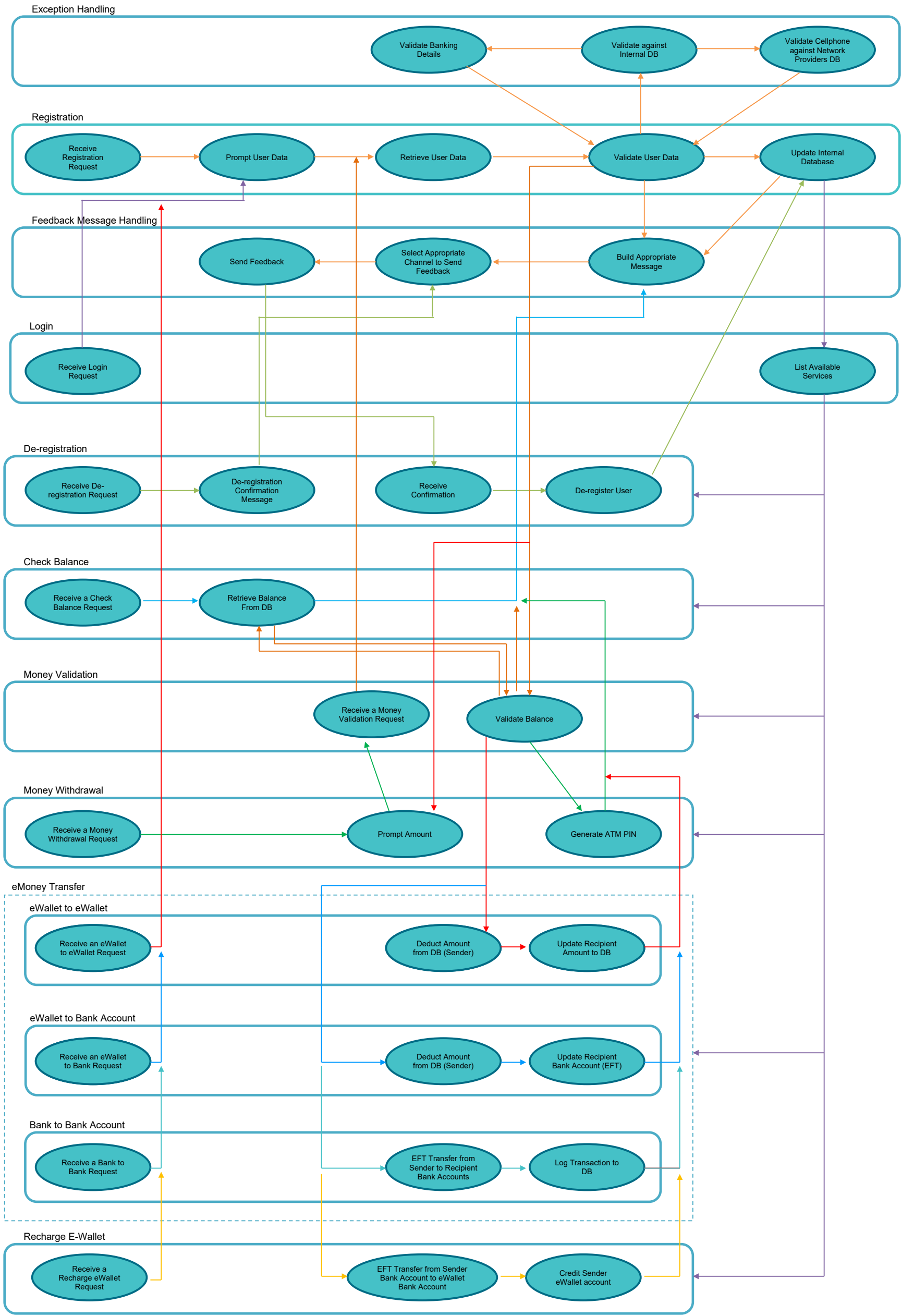

To get an understanding of the above Service model, let's follow the execution of some of the processes:

### A. REGISTRATION

Upon receiving a registration request, the system prompts the user for Data (Personal data, login details, banking details, Cellphone number…). The system then retrieves the appropriate data before validation of the data.

To validate the data, the system request services from the exception handling procedure that validate the data against the eWallet system internal database, Banking details and cellphone numbers against providers databases.

If the data are valid and no exception has been encountered, the system updates the internal database (new entry) and borrows services from the feedback message handling procedure to send a registration successful message to the new user.

Otherwise, an appropriate unsuccessful message is sent (Using feedback message handling services)

### B. LOGIN

Upon receiving a login request, the system borrows services from the registration to prompts the user for Data (login details …), to retrieve data, and to validate the data

If validation are correct, the system add logs to the DB and returns to the login procedure to list the available services

Otherwise, an appropriate unsuccessful message is sent (Using feedback message handling services)

### C. DE – REGISTRATION

Upon receiving a de-registration request, the system builds a confirmation message, borrows services from the feedback message handling to send the message

If the user validate the de-registration request, the system execute the de-registration (update database from the registration procedure) and send an appropriate message to the user via the feedback message handling services

### D. E-WALLET TO BANK ACCOUNT TRANSFER

Upon receiving an eWallet to Bank Account transfer request, the system borrows services from Registration to prompt user data and to validate the retrieved data. The system borrows services from money validation to validate the amount. If no exceptions are encountered, the system deducts the amount from the sender eWallet account and updates the receiver bank account via EFT. Finally, the system use the feedback message handling services to send a feedback to the sender.

## 8. CONCLUSION

Nowadays, an eWallet system provides an effective alternative to cashless transactions for both banked and unbanked population. Thanks to eWallet, financial transactions can successfully occur between those that have bank accounts and those that have bank accounts and those that do not have bank accounts.

In this paper, we have described a problem that can be solved by implementing a eWallet system. Mr. Kayembe Ka Tshitupa and his family can exchange money without constraints thanks to the system, as long as they have cellular phones at a minimum.

To show the implementation of the solution, the paper explores a series of business processes that translate into services that the system is expected to support. A technical architecture model as well as a service model is also described to give an idea of how the system can successfully be implemented.